\documentclass[]{interact}
\usepackage{caption}
\usepackage{amsmath,amssymb,amsfonts}
\usepackage[utf8]{inputenc}
\usepackage[english]{babel}
\usepackage{txfonts}
\usepackage{mathdots}
\usepackage{graphicx}
\usepackage{epsfig}
\usepackage{algorithm}
\usepackage[noend]{algpseudocode}
\usepackage{tabularray}
\usepackage{tikz}
\usepackage{lipsum}
\usepackage{booktabs}
\usepackage{url}

\usepackage[natbibapa,nodoi]{apacite}
\usepackage[bookmarks=true,colorlinks,linkcolor=black]{hyperref}
\usepackage{epstopdf}
\usepackage[caption=false]{subfig}

\theoremstyle{plain}

\theoremstyle{definition}

\theoremstyle{remark}

\begin{document}

\articletype{Research Paper}

\title{OrthoSeisnet: Seismic Inversion through Orthogonal Multi-scale Frequency Domain U-Net for Geophysical Exploration}

\author{
\name{Supriyo Chakraborty \textsuperscript{a}\thanks{CONTACT Supriyo Chakraborty. Email: meetsupriyochakraborty@gmail.com}, Aurobinda Routray\textsuperscript{b},
Sanjay Bhargav Dharavath\textsuperscript{c}
, and Tanmoy Dam\textsuperscript{d}}
\affil{\textsuperscript{a}Dept. ATDC, IIT Kharagpur, Kharagpur, India; \textsuperscript{b}Dept. of Electrical Engineering, IIT Kharagpur, Kharagpur, India
\textsuperscript{c}Dept. Physics, IIT Kharagpur, Kharagpur, India India
\textsuperscript{d}Dept. MAE, NTU Singapore, Singapore}
}

\maketitle

\begin{abstract}
Seismic inversion is crucial in hydrocarbon exploration, particularly for detecting hydrocarbons in thin layers. However, the detection of sparse thin layers within seismic datasets presents a significant challenge due to the ill-posed nature and poor non-linearity of the problem. While data-driven deep learning algorithms have shown promise, effectively addressing sparsity remains a critical area for improvement. To overcome this limitation, we propose OrthoSeisnet, a novel technique that integrates a multi-scale frequency domain transform within the U-Net framework. OrthoSeisnet aims to enhance the interpretability and resolution of seismic images, enabling the identification and utilization of sparse frequency components associated with hydrocarbon-bearing layers. By leveraging orthogonal basis functions and decoupling frequency components, OrthoSeisnet effectively improves data sparsity. 	We evaluate the performance of OrthoSeisnet using synthetic and real datasets obtained from the Krishna-Godavari basin. Orthoseisnet outperforms the traditional method through extensive performance analysis utilizing commonly used measures, such as mean absolute error (MAE), mean squared error (MSE), and structural similarity index (SSIM) \url{https://github.com/supriyo100/Orthoseisnet}.
\end{abstract}


\section{Introduction}
{S}{eismic} inversion is a pivotal technique in exploring oil and gas reservoirs, providing a detailed model of the subsurface. By employing sophisticated algorithms, seismic inversion estimates reflectivity based on seismic trace data, offering vital insights into geophysical formations. These insights enable direct forecasts of lithology, porosity, and other hydrocarbon indications, enhancing our understanding of subsurface formation dynamics \cite{sen2006seismic}. \par
 
 Seismic traces, representing acoustic waves propagating through the subsurface, are influenced by the medium's absorption, reflection, and refraction properties as shown by \cite{russell1988introduction}. In the study of sub-strata geology, seismic traces are inverted to acoustic impedance. However, this process encounters challenges due to the ill-posedness, non-uniqueness, and band-limited nature of seismic trace data. Traditional methods, including deconvolution and regularization techniques like Tikhonov regularization and total variation regularization (TV), have been employed to address these challenges in  \cite{shuey1985simplification, gholami2015nonlinear}.

A multifaceted approach has emerged to tackle inversion challenges in seismic data analysis. Diverse strategies, including regularization-based techniques, dynamic warping, and multi-scale inversion methods in \cite{burstedde2009algorithmic, gregor2010learning, lin2013ultrasound, bunks1995multiscale,chakraborty2020multi,chakraborty2022identification}, play a crucial role in mitigating the ill-posedness of inverse problems. Additionally, geostatistical seismic inversion methods, utilizing stochastic sequential simulation and co-simulation for model perturbation and updates have addressed the challenges associated with seismic data. \par

Integrating deep learning (DL) into seismic data analysis represents a transformative shift in this field. DL, encompassing deep neural networks (DNNs) and convolutional neural networks (CNNs), showcases remarkable proficiency in handling spatially structured multidimensional arrays. This versatility extends across various applications, including seismic inversion, velocity analysis, and seismic facies classification by \cite{alfarraj2019semisupervised, biswas2019prestack, feng2020unsupervised, jin2019dunet, mosser2018rapid, di2021seismic, liu2021deep, li2021consecutively}. Notably, Das et al. (2019) employed a 1D CNN for seismic impedance inversion \cite{das2019convolutional}.

However, the advantages of DL are accompanied by challenges, primarily stemming from the insufficiency of labelled data. To overcome this hurdle, strategies such as transfer learning, data augmentation, and synthetic data generation have been employed, underscoring the ongoing efforts to address data scarcity in seismic data analysis \cite{mosser2018rapid, das2019convolutional, sun2021physics}. Notably, the introduction of physics-driven neural networks by Karpatne et al. (2017) seek to enhance model-data relationships by leveraging unlabeled data \cite{karpatne2018machine}. \par

Introduction of neural network methods, such as recurrent neural networks (RNNs) and 1D CNNs, have demonstrated success in cases constrained by well logs \cite{jiang2022convolutional, tao2022seismic, he2022swin, cui2022map, gadylshin2022numerical, das2019convolutional, wang2022physics, gao2022global, wu2021deep, yuan2019impedance, wu2020seismic, sun2021physics}.

Addressing challenges specific to seismic impedance inversion, a novel approach involves the introduction of a self-attention U-Net, leveraging the noise resistance inherent in CNNs \cite{tao2023acoustic}. Persisting data scarcity, especially in well logs, necessitates innovative techniques such as data augmentation and semi-supervised learning \cite{ding2017augmentation, chen2018semi}. Physics-guided CNNs, incorporating wave-propagation physics during training, have been introduced for pre-stack and post-stack inversion in  \cite{biswas2019prestack}. Furthermore, adopting a closed-loop CNN structure, alternating between inversion and forward modeling, showcases the potential of deep learning to overcome traditional limitations in  \cite{wang2019computational}.

The application of DL in seismic inversion extends across acoustic impedance inversion, pre-stack elastic and lithological parameter inversion, and full waveform inversion by \cite{alfarraj2019semisupervised, das2019convolutional, wu2020seismic}. 2D network-based inversion methods, including U-Net, Attention Gate (AG), and unstructured convolutional neural network (UCNN), have emerged as an effective strategy to address spatial correlations, contributing to the development of more robust and efficient seismic inversion within the realm of deep learning \cite{ronneberger2015u, Cao2022, ning2023multichannel}. Interpretation of seismic data includes post-stack seismic inversion (PSTM) to resolve a clear picture of possible hydrocarbon deposition. Besides data scarcity, isolating thin layers is a challenge, as addressed above.  \par

Deep learning (DL) holds significant promise for seismic impedance inversion but grapples with challenges related to limited training data and the identification of thin layers. 2-D networks emerge as a solution, leveraging spatial continuity to address data scarcity and incorporating sparsity for more effective thin-layer identification. The sparsity property reduces computational complexity and enhances model performance since thin layers are sporadically distributed within the overall geological structure. The presented paper aims to overcome these challenges, recognizing that identifying thin layers is pivotal in improving inversion accuracy and boosting success rates in hydrocarbon exploration.

In thin layer detection, the orthogonality inherent in the UNet architecture represents sparse layers within the seismic trace. The paper introduces a novel approach using multi-scale 2D frequency domain transformation for multi-trace inversion to achieve orthogonality. Applying the inverse to the frequency domain representations of each layer in the UNet makes the resulting time domain representations sparse, with most values being zero. This intentional sparsity reduces the model's computational complexity, enhancing its accuracy and efficiency. The UNet architecture, complemented by techniques like early stopping, skip connections with multi-scale 2D frequency domain transformation, and normalization, facilitates the efficient estimation of sparse layers from seismic trace images. The methodological contributions in this paper can be outlined as follows:

\begin{enumerate}
    \item Introducing a novel approach that employs multi-scale 2D frequency domain transformation within the UNet architecture, facilitating an enhanced sparsity representation.
    \item Incorporating multi-trace inversion using a 2D CNN within the UNet block, ensuring lateral continuity among traces is preserved.
    \item Conducting extensive synthetic and real data experiments to validate the proposed algorithms.
\end{enumerate}
In Section II, we review related works in the field of Deep Learning (DL) applied to PSTM (Pre-Stack Time Migration) seismic data. The methodology is presented in Section III, followed by experimental results in Section IV. Section V contains the discussion, and the conclusion is presented in Section VI.

\section{Background}

Recent seismic inversion advances are attributed to applying deep learning (DL) techniques, recognized for capturing intricate nonlinear relationships in seismic data~\cite{lecun2015deep}. DL, distinguished from traditional model-driven methods, utilizes adjustable parameters for learning from training datasets and exhibits remarkable performance in diverse domains~\cite{lecun2015deep}.

DL's success extends to geophysical applications, showcasing potential in learning geological formations~\cite{sun2020theory} and fault detection~\cite{wu2019faultnet3d}. Convolutional Neural Networks (CNNs) \cite{dam2021improving, dam2022latent} demonstrate promise in predicting acoustic impedance (AI), which is pivotal for reservoir characterization. However, one-dimensional CNNs face challenges in real data due to poor lateral continuity~\cite{das2019convolutional}. To address this, a physics-guided CNN tailored for pre-stack inversion has been proposed~\cite{biswas2019prestack}.

DL-based seismic inversion encounters a challenge in the need for substantial labeled data. Mitigating this, semi-supervised CNNs leverage labelled and unlabeled data, improving results by utilizing abundant unlabeled seismic data~\cite{alfarraj2019semisupervised}. Domain adaptation networks, incorporating unlabeled data during training, enhance performance~\cite{wang2022seismic}.

Studies contribute to DL advancements in seismic inversion, including a semi-supervised sequence modelling approach for elastic impedance inversion~\cite{alfarraj2019semisupervised} and InversionNet, a real-time full-waveform inversion method incorporating CNNs and continuous conditional random fields (CRFs) for improved efficiency and accuracy~\cite{wu2018inversionnet}. Comparative studies provide insights into different DL methods for seismic impedance inversion~\cite{zhang2022comparison}. A deep-learning inversion framework emphasizes its capability to learn complex features and enhance accuracy~\cite{li2019deep}.
Exploration extends to three-dimensional (3D) approaches in seismic inversion, exemplified by 3D full-waveform inversion in the time-frequency domain applied to field data~\cite{tran20203d}. An encoder-decoder architecture efficiently handles large-scale 3D seismic inversion~\cite{gelboim2022encoder}.

DL techniques are harnessed in specific applications, such as salt body classification for salt interpretation and reservoir characterization~\cite{ul2020using}. An autoencoder-based model introduces dimensionality reduction for large-dimensional seismic inversion, enhancing efficiency while preserving accuracy~\cite{gao2020large}. State-of-the-art works introduce novel techniques, including a seismic inversion method based on two-dimensional CNNs and domain adaptation to tackle domain mismatch between training and testing data~\cite{wang2022seismic}. A multichannel seismic impedance inversion approach based on Attention U-Net incorporates attention mechanisms to capture important features and improve accuracy~\cite{ning2023multichannel}.
These studies demonstrate the potential of DL techniques in seismic inversion, showcasing improvements in inversion accuracy, computational efficiency, and the ability to handle complex subsurface structures. DL-based seismic inversion can capture complex nonlinear relationships, allowing for more accurate and reliable inversion results compared to traditional model-driven methods. Current literature has yet to target sparseness. Specifically, we have introduced an orthogonal seismic inversion network (Orthoseisnet) that uses the Unet backbone with multi-scale frequency domain transformation to target thin layers for higher resolution in PSTM seismic data.

\section{Methodology}
This section discusses seismic inversion theory, showing challenges in inversion and traditional ways to estimate optimized solutions. The architecture modification of the Unet backbone with a multi-scale frequency domain using FFT and IFFT is discussed in detail.
\subsection{Theory}
	A seismic impedance inversion is a foremost tool for exploring hydrocarbon in seismic exploration. By convention, we can define forward and inverse problems as: \begin{itemize}
		\item \textbf{Forward problem:} Given a model of the subsurface, generate seismic traces.
		\item \textbf{Inverse Problem: } Inverse problem: Given seismic traces, estimate the subsurface model.
	\end{itemize}	
	\subsubsection{Forward problem}
	The forward model for seismic trace generation can be represented as:
	\begin{equation}
		S(t)=w(t)\ast r(t) + \in(t)
	\end{equation}
	
	where $S(t)$ is the seismic trace, $w(t)$ is the wavelet, $r(t)$ is the reflectivity, and $\in(t)$ is the noise. The wavelet gives the shape of the seismic signal. At the same time, reflectivity represents the changes in acoustic impedance in the sub-surface, whereas noise is the random function representing the errors in the dataset. The reflectivity can be calculated using the following equation:
	
	\begin{equation}
		r(t) = \frac{{di(t_{p})}}{{dt}}\frac{1}{{2i_p(t)}} = \frac{{\Delta \ln (i{p_i})}}{2}
	\end{equation} Where  the acoustic primary impedance$(ip)$ is given by:
	\begin{equation}
		i_{p} = \nu*\rho
	\end{equation}
 where $\nu$ is primary velocity and $\rho$ is the density.
	\subsubsection{Inverse problem}
	For the inverse problem, we estimate reflectivity from a given seismic trace data considering a dictionary of wavelets. Due to its non-unique and under-determined nature, inversion is ill-posed. To make the inverse problem well-posed, we need to add regularization terms to the objective function. Regularization terms penalize solutions inconsistent with prior knowledge or assumptions about the subsurface. The objective function minimizes the error by optimizing the model based on converging criteria, resulting in a stable solution:
	\begin{equation}
		\phi (m) = \left\| {d - Gm} \right\|_2^2 + \chi {\left\| m \right\|_1}
	\end{equation}
	Where the objective function is $\phi (m) $, $d$ is the observed seismic data, $G$ is the measurement matrix, $m$ is the model reflectivity, and  $\chi$ is a regularization term. The choice of regularization term depends on the desired solution properties, such as $L_{1}$ norm, total variation, or smoothness constraints. The solution to the optimization problem is the estimated model parameters $\mathbf{m}^{\star}$ that best fit the observed data and satisfy the regularization constraints.

	\subsection{\textbf{Network Architecture}}	
\begin{figure*}
\centering
{\includegraphics[scale=0.36]{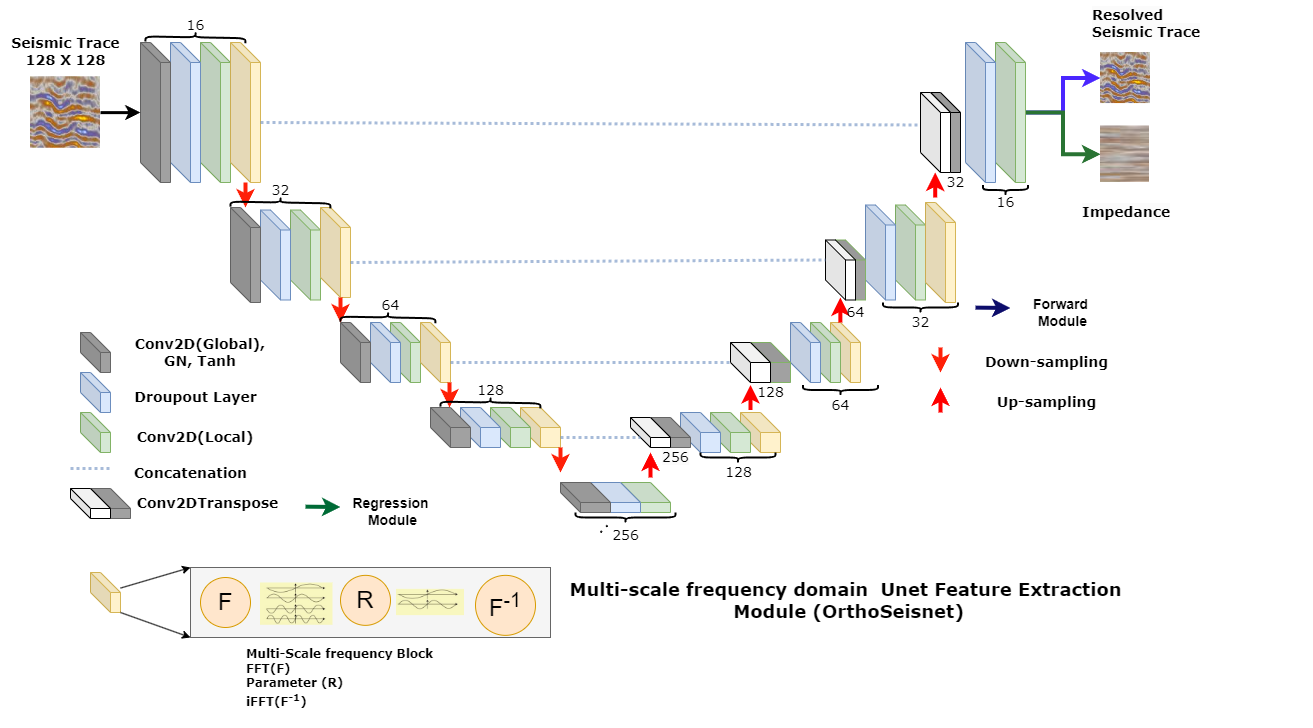}}
\caption{Feature Extraction Module for Multi-scale Frequency domain Unet} 
\label{fig: Multiscale Frequency domain Unet } 
\end{figure*}

    This paper uses data-driven methods to solve the reflectivity inversion problem using Unet for feature learning and extraction. The network maps seismic trace to impedance, and algorithm output behaves like inversion but is constrained solely by data. The representation ability of DL is used as black boxes as it helps to predict the network relationship between seismic trace and impedance. The network architecture described above follows a U-Net-like structure commonly used in image segmentation tasks. The architecture consists of a contraction and expansive paths, as shown in Figure ~\ref{fig: Multiscale Frequency domain Unet }.	In Fourier-DeepONet ~\cite{zhu2023fourier}, Fourier, along with Unet, was used to create architecture, but the network has several Unet architectures, increasing complexity and possibility of over-fitting. Our work aims to keep the network light while resolving high and low-frequency features. Figure ~\ref{fig: Multiscale Frequency domain Unet } shows Orthoseisnet architecture, which orthogonalizes frequency domain components to recover sparse features. The FFT-iFFT block includes R, which is the parameterization in Fourier space.
    
	The model is compiled using the Adam optimizer, mean squared error (MSE), SSIM and other losses. The accuracy metric is also calculated. Unet input $\mathbf{X} $ and output $\mathbf{Y}$ can be shown as below:	
	\begin{equation}
		\mathbf{X, Y} \in \mathbb{R}^{H \times W \times C}
	\end{equation}
 Where H, W, and C are the feature graph's height, width, and number of channels, respectively. The overall data flow is shown in Figure ~\ref{fig: Multiscale Frequency domain Unet } details, which are given in architecture Table ~\ref{tab: Orthoseisnet Network Architecture}.
\subsubsection{\textbf{Encoder Layer}}
 
The convolutional Layer is of Kernel size: $3 \times 3$ starting with filters: $16$, using activation function: $\tanh$. Output tensor: $\mathbf{Y^{L}} \in \mathbb{R}^{H \times W \times 16}$. The height $H$ and width $W$, is processed through the following operations:

\begin{equation}
    Y_m = \mathcal{}{m}(\tanh \left( \text{Conv2D}(\mathcal{D} \left\{ \tanh \left( \text{Conv2D}(X, K) \right) \right\}, K') \right) )
\end{equation}

\begin{equation}
    \mathbb{f} = \mathcal{F}^{-1} \left\{ \mathcal{F}(Y_m) \right\}
\end{equation}
where maxpooling $\mathcal{}{m}$ dropout $\mathcal{D} (0.1-0.3)$, and  convolution is given by :
	\begin{equation}
		Y^{L}1_{i,j,k} = \tanh\left(\sum_{k'=0}^{2} \sum_{l'=0}^{2} X_{2i+k', 2j+l', 1} \cdot K_{k',l',1,k}\right)
	\end{equation}

\subsubsection{Multi-scale Frequency Domain Transform Layer}:
Certainly! Here are the LaTeX equations for the provided descriptions:
   
A) 2D Fourier transform $\mathbb(F)$is applied to the local 2D Convolution layer output.
	 Input: $\mathbf{Y}$, Output: $\mathcal{F} \in \mathbb{C}^{\frac{H}{2} \times \frac{W}{2} \times 16}$ (complex-valued)
	\begin{equation}
		\mathcal{F}_{u,v,k} = \frac{1}{\frac{H}{2} \cdot \frac{W}{2}}\sum_{x=0}^{\frac{H}{2}-1}\sum_{y=0}^{\frac{W}{2}-1}Yp_{x,y,k}\exp\left(-j2\pi\left(\frac{ux}{\frac{H}{2}}+\frac{vy}{\frac{W}{2}}\right)\right)
	\end{equation}
 Where $u=0,1,...,\frac{H}{2}$ and  $v = u=0,1,...,\frac{W}{2}$ 
 
 B) Parameterizing weight tensor (R) with \(\kappa\) directly by its Fourier coefficients:
\begin{equation}\left(\mathcal{K}\left(v_l\right)\right)(x) = \mathcal{F}^{-1}\left(R \cdot \mathcal{F}\left(v_l\right)\right)(x), \quad \forall_x \in D \end{equation}
where \( \mathcal{F} \) denotes the Fourier transform of a function \( f: D \rightarrow \mathbb{R}^c \) and \( \mathcal{F}^{-1} \) is its inverse.
C)Expressing multiplication by the learnable weight tensor \( R \):
\begin{equation} \left(R \cdot \mathcal{F}\left(v_l\right)\right)_{k, i} = \sum_{j=1}^c R_{k, i, j} \left(\mathcal{F}\left(v_l\right)\right)_{k, j}, \quad \forall k=1..k_{\max }, i=1..c \end{equation}
D)Followed by  Inverse 2D Multi-scale Frequency Domain Transform Layer $\mathbb{F^{-1}}$ :
	Input: $\mathcal{F}$
	- Output: $\mathbb{f} \in \mathbb{C}^{\frac{H}{2} \times \frac{W}{2} \times 16}$ (complex-valued)
	\begin{equation}
		\mathbb{f}_{x,y,k} = \sum_{u=0}^{\frac{H}{2}-1}\sum_{v=0}^{\frac{W}{2}-1}\mathcal{F}_{u,v,k}\exp\left(j2\pi\left(\frac{ux}{\frac{H}{2}}+\frac{vy}{\frac{W}{2}}\right)\right)
	\end{equation}
	The output of $1^{st}$ Unet encoding layer is fed to the next layer. Where the complex-valued output of the previous layer is used as the absolute value input to the first convolutional layer in the next encoding layer. The output tensor of this convolutional layer, denoted as $\mathbf{YL1}$, is computed as:
	\begin{equation}
		\mathbf{YL1}{i,j,k} = \tanh \left(conv2D(Y_m,K)\right)\left|\mathbb{f}\right|
	\end{equation}
	Where $\mathbf{YL1}$ represents the output tensor of the first convolutional layer in the next encoding layer, the last encoding layer consists of two conv2d layers (activation='tanh', kernel size=$3\times3$) with a dropout layer between them.
	
	\subsubsection{\textbf{Decoder Layer}}
\begin{equation}
    {x^{i + 1}} = \tanh \left\{ \mathcal{D} \left\{ \text{ReLU} \left( \text{Conv2D} \left( \mathbb{C} \left\{ \text{Conv2D} \left( \mathbb{Z}L2_{i,j,k} \cdot K \right) \right\} \right) \right) \right\} \right\}
\end{equation}

	Where Concatenation is $\mathbb{C}$ and $\mathbb{Z}L2_{i,j,k}$ is given by
	\begin{equation}
		\mathbb{Z}L2_{i,j,k} = \left( \mathbb{C} \left\{ \text{Conv2D} \left( {W^t} \cdot {x_i} \right) \right\} \right)
	\end{equation}
	ReLU activation is used with a Convolution-2D layer with a dropout $\mathbb{D}$ layer with $0.1 to 0.3$. The Multi-scale Frequency Domain Transform Layer gives the final layer for every decoder layer:
	\begin{equation}
		\mathcal{F}_{u,v,k} = \frac{1}{\frac{H}{2} \cdot \frac{W}{2}}\sum_{x=0}^{\frac{H}{2}-1}\sum_{y=0}^{\frac{W}{2}-1}\mathbb{Z}L2_{x,y,k}\exp\left(-j2\pi\left(\frac{ux}{\frac{H}{2}}+\frac{vy}{\frac{W}{2}}\right)\right)
	\end{equation}
	
	Inverse 2D Multi-scale Frequency Domain Transform Layer:
	\begin{equation}
		\mathbb{f}_{x,y,k} = \sum_{u=0}^{\frac{H}{2}-1}\sum_{v=0}^{\frac{W}{2}-1}\mathcal{F}_{u,v,k}\exp\left(j2\pi\left(\frac{ux}{\frac{H}{2}}+\frac{vy}{\frac{W}{2}}\right)\right)
	\end{equation}
		The absolute value from the Multi-scale Frequency layer is used as input to the next decoder layer as 	
	\begin{equation}
		\mathbb{Z}L1^{'}_{i,j,k} = \tanh \left\{ \text{Conv2D} \left\{ Z \cdot K \right\} \left| \mathbb{f}_{x,y,k} \right| \right\}
	\end{equation}
 The final layer uses soft-max to output the estimated resolved image. \par

 \begin{table}[!ht]
    \centering
    \caption{Orthoseisnet Network Architecture}
    \label{tab: Orthoseisnet Network Architecture}
    \begin{tabular}{|c|c|c|c|}
        \hline
        \textbf{Block} & \textbf{Layer} & \textbf{Unit} & \textbf{Parameters} \\ \hline \hline
        Input & 0 & Seismic 2D ($128\times128\times1$) & 0 \\ \hline
        Enc1 & 1 & Conv2D(16, (3,3),tanh)+GN+L1 & 160+32\\ \hline
        ~ & 2 & Dropout, Conv2D+GN & 2320+32 \\ \hline
        ~ & 3-4 & MaxPooling, FFT, iFFT & 0  \\ \hline
        Enc2 & 5-8 & Enc(32) abs(iFFT) & 14016 \\ \hline
        Enc3 & 9-12 & Enc(64) abs(iFFT) & 55680 \\ \hline
        Enc4 & 13-16 & Enc(128) abs(iFFT) & 221952 \\ \hline
        Enc5 & 17 & Conv2D(256) abs(iFFT) & 295168\\ \hline
        Dec1 & 18 & Concat & 1.312M \\ \hline
        ~ & 19 & Conv2DTrans(128)+GN & 110976 \\ \hline
        ~ & 20 & Dropout, Conv2D & 147584 \\ \hline
        ~ & 21 & FFT, iFFT & ~ \\ \hline
        Dec2 & 22-26 & Dec(64), Dec(32), Dec(16) & 37k, 2320 \\ \hline
        Dec3 & 22-26 & Dec(64), Dec(32), Dec(16) & 37k, 2320\\ \hline
        Dec4 & 22-26 & Dec(64), Dec(32), Dec(16) & 37k, 2320 \\ \hline
        ~ & ~ & Total Parameters & 1,940,817 \\ \hline \hline
    \end{tabular}
\end{table}
26816
This architecture  Table ~\ref{tab: Orthoseisnet Network Architecture} is designed for processing seismic data, leveraging conventional convolutional layers and operations in the frequency domain (FFT/iFFT). The table  Table ~\ref{tab: Orthoseisnet Network Architecture} offers insights into the structure and complexity of the Orthoseisnet neural network.

\subsection{Loss function}

The impact of the loss function, network architecture, and data type on the inversion resolution of complex stratigraphic boundaries has been thoroughly investigated \cite{li2019deep, liu2021deep, wu2019inversionnet, araya2019fast}. A novel approach combining the structural similarity index (SSIM) and L1 norm was proposed to improve the accuracy of extracting construction boundaries and geological structures. Incorporating SSIM into the loss function significantly enhances the inversion resolution~\cite{li2019deep}.

The proposed approach utilizes a mixed loss function that combines mean squared error (MSE) and SSIM to optimize pixel and local patch velocity misfits simultaneously. Synthetic and real data were used to validate this approach, which outperformed using only MSE as the loss function.

In conclusion, this study thoroughly investigates the impact of the loss function, network architecture, and data type on the inversion resolution of complex stratigraphic boundaries. The proposed approach, which integrates SSIM into the loss function, demonstrates its effectiveness in accurately extracting geological structures and improving the inversion resolution.

Different loss functions, including MSE, Mean Absolute Error (MAE), and SSIM, were considered in the U-Net architecture for pixel-wise evaluation. Ms-SSIE incorporates multi-scale analysis, improving image structure sensitivity, scaling, and translation robustness, and higher correlation with human perception of image quality~\cite{liu2021deep}.

Mean Absolute Error (MAE) measures the average absolute difference between a dataset's predicted and actual values. It is a robust metric that indicates the model's prediction accuracy~\cite{wu2019inversionnet}.

For a dataset with $N$ samples, the MAE is calculated as:

\begin{equation}
    \text{MAE} = \frac{1}{N} \sum_{i=1}^{N} | \hat{y}_i - y_i |
\end{equation}

Here, $\hat{y}_i$ represents the predicted value for the $i$-th sample, and $y_i$ represents the corresponding actual value. The absolute differences between predicted and actual values are summed and then averaged across all samples to obtain the MAE.

The MSE loss is represented as ~\cite{wu2019inversionnet}:

\begin{equation}
    \mathcal{L}_{\text{MSE}} = \frac{1}{N} \sum_{i=1}^{N} \| \hat{y}_i - y_i \|^2 
\end{equation}

The Structural Similarity Index (SSIM) measures the similarity between two images, considering contrast and structure. The SSIM index produces a value between -1 and 1, where 1 indicates perfect similarity. SSIM loss for $I(i, j), K(i, j)$ ~\cite{araya2019fast} is represented as SSIM in the equation below:

\begin{equation}
    SSIM = \frac{{(2 \cdot \mu_{I(i, j)} \cdot \mu_{K(i, j)} + C_1) \cdot (2 \cdot \sigma_{I(i, j), K(i, j)} + C_2)}}{{(\mu_{I(i, j)}^2 + \mu_{K(i, j)}^2 + C_1) \cdot (\sigma_{I(i, j)}^2 + \sigma_{K(i, j)}^2 + C_2)}}
\end{equation}
Where $I(i, j)$ and $K(i, j)$ are the pixel values of the two images being compared, \(\mu_{I(i, j)}\) and \(\mu_{K(i, j)}\) are the means of the pixel values in the local windows of images \(I\) and \(K\) and \(\sigma_{I(i, j), K(i, j)}\) is the covariance of the pixel values in the local windows of images \(I\) and \(K\). \par
\(\sigma_{I(i, j)}^2\) and \(\sigma_{K(i, j)}^2\) are the variances of pixel values in the local windows of images \(I\) and \(K\). \(C_1\) and \(C_2\) are constants added to avoid instability when the denominators are near zero.
The advantage of using SSIM in seismic inversion, particularly in the context of a U-Net-based model, lies in its ability to capture perceptual differences between images, unlike traditional pixel-wise metrics such as MSE.

\section{Experiment}
The proposed OrthoSeisnet network is validated on synthetic
and real seismic trace data.

\subsection{\textit{Synthetic data}}	
\begin{figure}[!ht]
	\centering
	{\includegraphics[scale=0.8]{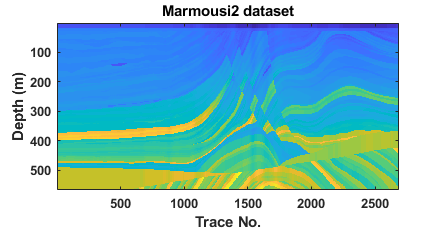}}
	\caption{ 2D Marmousi2 Synthetic Dataset}
	\label{fig:Marmousi2}
\end{figure}

We conducted experiments using the Marmousi synthetic dataset, known for its geological complexity Figure ~\ref{fig:Marmousi2}. The dataset comprises 13,200 traces with 1001 samples per trace at a 1 ms time interval. Experiments were implemented in Python 3.6 using TensorFlow on Google Colab TPUs. The network, OrthoSeisnet, had input/output sizes of $128\times128$, an initial learning rate of $1\times10^{-2}$, and was trained for 100 epochs. OrthoSeisnet has 1.9 million parameters, with pre-training taking 0.9 min/sample, fine-tuning taking 0.8 min, and an inference time of 0.09 seconds.

OrthoSeisnet consistently outperformed existing methods with increased training data. Noisy data experiments at signal-to-noise ratios ($\mathcal{SNR}$) of $10$ dB, $20$ dB, and $30$ dB for elastic parameter inversion ($V_p$, $V_s$, $\rho$) favored OrthoSeisnet over model-driven approaches. MSE values confirmed the superiority of data-driven methods, with OrthoSeisnet exhibiting the smallest MSE.
In conclusion, despite increased computational demands, the Marmousi synthetic dataset served as an effective testing ground for data-driven models, particularly OrthoSeisnet, showcasing their potential in seismic inversion tasks.

 \begin{figure}[!ht]
	\centering
	{\includegraphics[scale=0.6]{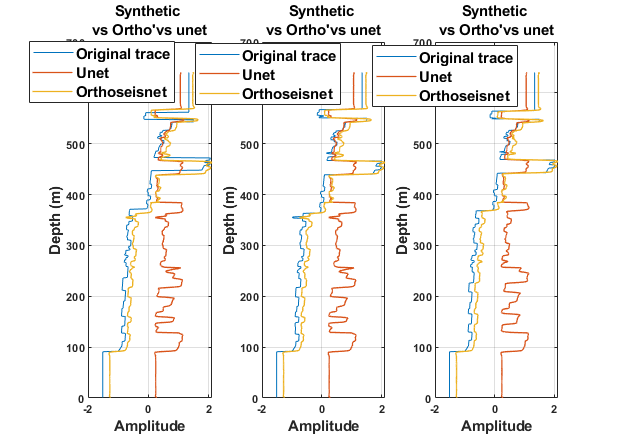}}
	\caption{ Marmousi2 Vs. Orthoseisnet Vs Unet Trace No 2, 400, 200}
	\label{fig:Marmousi2 Vs. Orthoseisnet Vs Unet}
\end{figure}
 \begin{figure}[!ht]
	\centering
	{\includegraphics[scale=0.6]{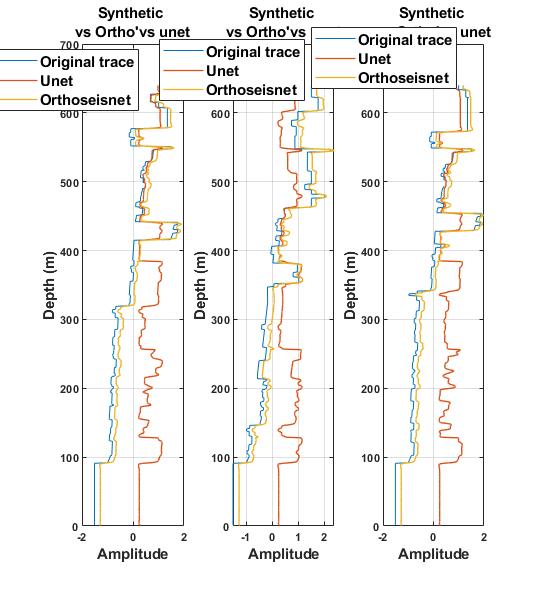}}
	\caption{  Marmousi2 Vs. Orthoseisnet Vs Unet Trace No 800, 1000, 2000}
	\label{fig:Marmousi2 Vs. Orthoseisnet Vs Unet}
\end{figure}
	
 	\begin{figure*}[!ht]
		\centering 
		\scalebox{0.4} 
		{\includegraphics{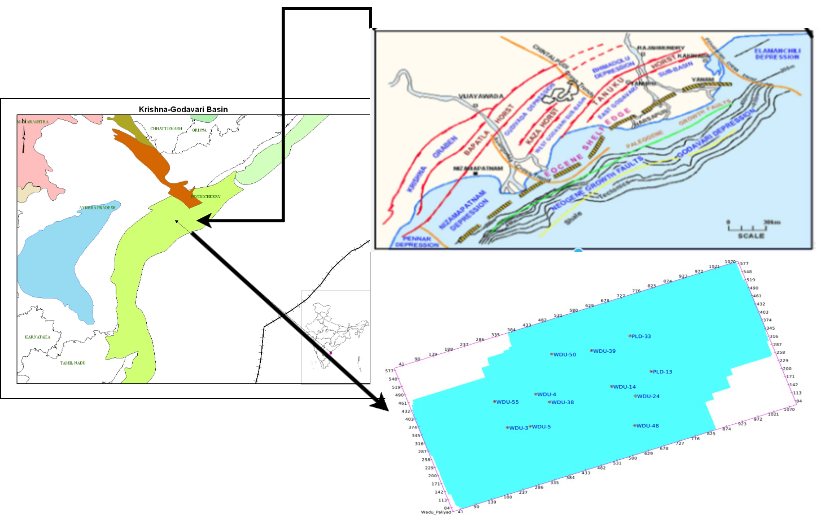}} 
		\caption{Krishna-Godavari basin Courtesy: National data repository, India} 
		\label{fig:kg-basin} 
	\end{figure*}
 \subsection{Real Data Experiment}
	The data provided by the Oil and Natural Gas Corporation of India(ONGC) is 3D Post-stack seismic trace data under the GEOPIC project. The data is $150 sq. km$ with $12$ wells in the Krishna-Godavari basin. This basin is an extensive deltaic plain formed by the Krishna and Godavari rivers in Andhra Pradesh and the adjoining areas of the Bay of Bengal, shown in Figure ~\ref{fig:kg-basin}. The basin is a proven petroliferous basin located on the east coast of India with sediments about 5 km thick ranging in age from Late Carboniferous to Pleistocene. The major geo-morphological units of the basin are upland plains, coastal plains, recent floods, and delta plains. Geological surveys, gravity-magnetic surveys, and seismic coverage have been carried out in the area. More than $225$ prospects have been probed by drilling over $557$ exploratory wells, and hydrocarbon accumulations have been proven in $75$ of these prospects. The basin's tectonic history comprises the rift, passive margin, compressional, and post-compressional stages. The data provided is shown in the table above. The survey area of the data can be seen if the Figure ~\ref{fig:kg-basin} Input size of $128\times 128$ is used just like synthetic data but across the Inline and Crosslines, as shown in the data below. 
\begin{table}[!ht]
    \centering
    \caption{KG Basin Dataset Details}
    \begin{tabular}{ll}
        \toprule
        \textbf{Parameter} & \textbf{Value} \\
        \midrule
        Inline Range & 70-585 \\
        Crossline Range & 17-1093 \\
        Bin Size & 17.5m$\times$17.5m \\
        Sampling Interval & 2ms \\
        Data Length & 0-2sec \\
        \bottomrule
    \end{tabular}
\end{table}

\begin{figure*}[!ht]
		\centering
            {\includegraphics[scale=0.58]{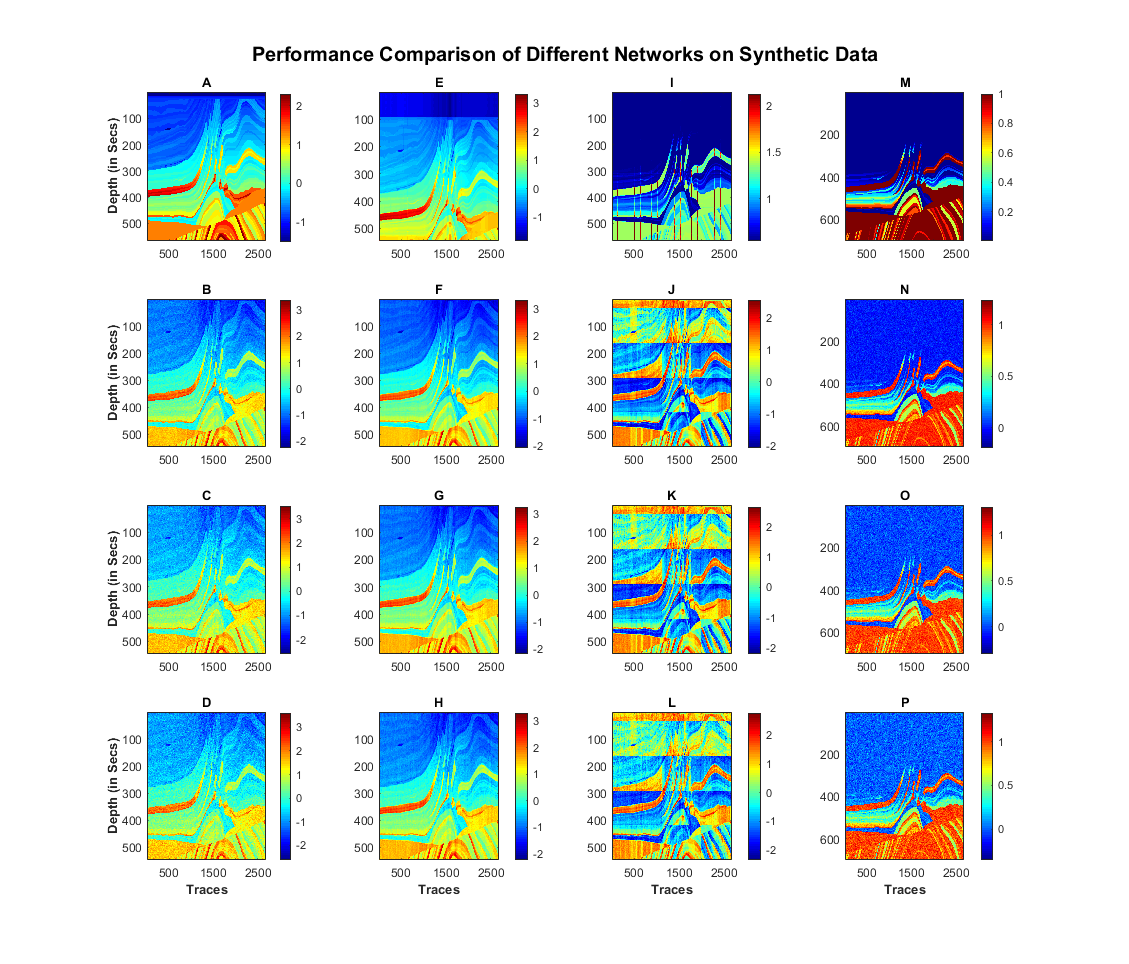}}
		\caption{(A-D) marmousi2 data 0dB, 10dB, 20dB, 30dB noise (E-H) Orthoseisnet 0dB, 10dB, 20dB, 30dB noise (I-L) Unet 0dB, 10dB, 20dB, 30dB 1D CNN 0dB, 10dB, 20dB, 30dB noise.}
		\label{fig:comparative-result}
\end{figure*}

\begin{table}
	\centering
	\caption{Table 1: Performance Comparison of Different Networks on Synthetic Noisy Dataset }
        \label{Table: Experiment on Synthetic Baseline Model}
	\begin{tblr}{
			width = \linewidth,
			colspec = {Q[200]Q[198]Q[113]Q[135]Q[113]},
			cell{2}{1} = {r=4}{},
			cell{6}{1} = {r=4}{},
			cell{10}{1} = {r=4}{},
			cell{14}{1} = {r=4}{},
			vlines,
			hline{1-2,6,10,14,18} = {-}{},
			hline{3-5,7-9,11-13,15-17} = {2-6}{},
                row{odd} = {bg=gray9},
		}
		{\textbf{Data-}\\\textbf{set}}     & \textbf{Network}    & \textbf{MAE}     & \textbf{MSE}      & \textbf{SSIM}  \\
		{\textbf{Synthe-}\\\textbf{tic (0db)}\\\textbf{noise}}  & OrthoSeisnet(ours)       & \textbf{0.043}  & \textbf{0.005}    & \textbf{0.991} \\
		                   & Inversion-net~\cite{wu2019inversionnet}      & \textit{0.95}    & \textit{1.261}    & \textit{0.846} \\
		                   & Unet~\cite{ronneberger2015u}              & 2.43             & 3.12              & 0.623          \\
		                   & CNN~\cite{alfarraj2019semisupervised}        & 2.65             & 3.22              & 0.636          \\
		{\textbf{Synthe-}\\\textbf{tic (10db)}\\\textbf{noise}} & OrthoSeisnet(ours)       & \textbf{0.083}  & \textbf{0.012}    & \textbf{0.855} \\
		                   & Inversion-net~\cite{wu2019inversionnet}      & \textit{0.861}   & \textit{1.0597}   & \textit{0.701} \\
		                   & Unet~\cite{ronneberger2015u}              & 2.74             & 3.331             & 0.602          \\
		                   & CNN~\cite{alfarraj2019semisupervised}        & 2.78             & 3.35              & 0.611          \\
		{\textbf{Synthe-}\\\textbf{tic (20db)}\\\textbf{noise}} & OrthoSeisnet(ours)       & \textbf{0.112}  & \textbf{0.021}    & \textbf{0.803} \\
		                   & Inversion-net~\cite{wu2019inversionnet}  & \textit{0.871}   & \textit{1.083}   & \textit{0.635} \\
		                   & Unet  ~\cite{ronneberger2015u}            & 3.15             & 3.56              & 0.542          \\
		                   & CNN~\cite{alfarraj2019semisupervised}        & 3.09             & 3.42              & 0.533          \\
		{\textbf{Synthe-}\\\textbf{tic (30db)}\\\textbf{noise}} & OrthoSeisnet(ours)      & \textbf{0.134}  & \textbf{0.032}    & \textbf{0.772} \\
		                   & Inversion-net~\cite{wu2019inversionnet}     & \textit{0.889}   & \textit{1.118}   & \textit{0.608} \\
		                   & Unet  ~\cite{ronneberger2015u}            & 3.52             & 3.89              & 0.511          \\
		                   & CNN~\cite{alfarraj2019semisupervised}        & 3.76             & 4.16              & 0.513
	\end{tblr}
\end{table}
	
	A comparative analysis of synthetic data at various noise levels is conducted for OrthoSeisnet, Inversionnet \cite{wu2018inversionnet}, and Basis Pursuit (BPI) \cite{zhang2011}. Table ~\ref{Table: Experiment on Synthetic Baseline Model} presents the performance of these networks on noisy synthetic datasets, with a specific focus on OrthoSeisnet. The discussion centers on the improvements observed in the loss metrics (MAE, MSE, SSIM) compared to Inversionnet and SeisInvnet. The table provides a comprehensive comparison across different noise levels (0db, 10db, 20db, 30db) and evaluates Mean Absolute Error (MAE), Mean Squared Error (MSE), and Structural Similarity Index (SSIM). OrthoSeisnet consistently outperforms Inversion-net, Unet, and SeisInvnet, showcasing lower MAE and MSE values and higher SSIM values. This emphasizes OrthoSeisnet's superior accuracy and robustness in handling synthetic noisy datasets, making it a promising choice for seismic data prediction. The results indicate that OrthoSeisnet effectively preserves structural patterns, as reflected in its high SSIM values approaching 1. The table offers a clear and concise overview of each network's performance metrics under different noise conditions.

	Figure ~\ref{fig:Marmousi2 Vs. Orthoseisnet Vs Unet}  compares OrthoSeisnet with the latest state-of-the-art algorithm concerning different noise levels in the Marmousi2 synthetic model. The clarity of layers from OrthoSeisnet shows the algorithm's effectiveness. Figure ~\ref{fig:comparative-result}. shows a bigger view of the synthetic model at 20db noise.

	It's important to note that the improvements mentioned above are specific to the respective metrics (MAE, MSE, SSIM). They are calculated by comparing the values of OrthoSeisnet to Inversionnet. For the noisy synthetic datasets with 10 dB, 20 dB, and 30 dB noise levels, you can perform similar calculations to determine the percentage improvement in the loss metrics for OrthoSeisnet compared to Inversionnet and SeisInvnet.
	
	In summary, OrthoSeisnet exhibits improvement in the loss metrics (MAE, MSE, SSIM) compared to Inversionnet and SeisInvnet on noisy synthetic datasets, with percentage improvements varying across the metrics.

	\subsection{Experiment on Real data}
	
\begin{figure*}[!ht]
    \centering
    \includegraphics[scale=0.6]{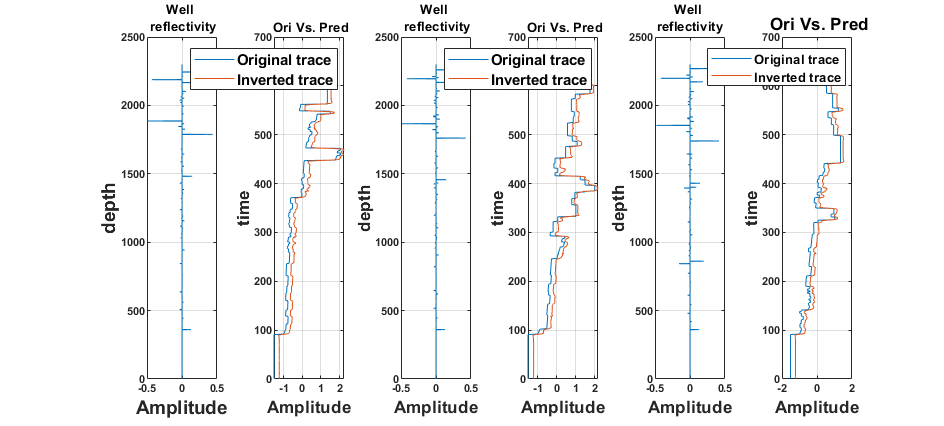}
    \caption{Original seismic data Vs. Orthoseisnet at Inline with neighbouring  Well 1-3 }
    \label{fig:comparative-result real well}
\end{figure*}

	OrthoSeisnet on real data improves the $figure$ and $Table 2$.
\begin{figure*}[!ht]
    \centering
    \includegraphics[scale=0.45]{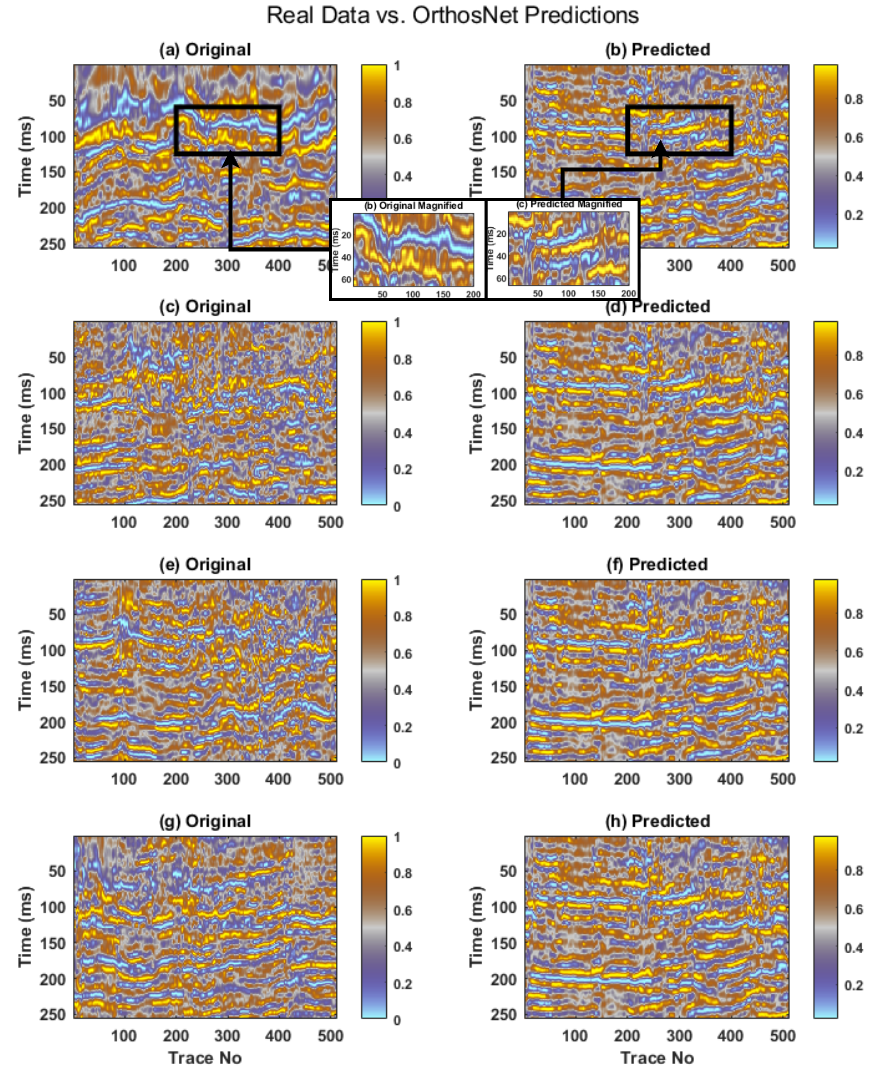}
    \caption{Comparison between Original Seismic Data and Orthoseisnet at Inline Sections: (a-b) Inline 39, (c-d) Inline 277, (e-f) Inline 316, and (g-h) Inline 460.}
    \label{fig:comparative-result}
\end{figure*}
Comparison with the well-log of KG basin well GS-29-10, DWN-S-1 and DWN-Q-1 and their neighbouring original seismic trace to estimated trace is shown in Figure ~\ref{fig:comparative-result real well}.

\begin{table}[!ht]
\centering
\caption{Performance Comparison of Different Networks on Real Data}
\label{Table:RealDataComparison}
\begin{tabular}{
l|l|l|l|l}
\hline
Network & MSE & MAE & SSIM & R2 \\ \hline
BPI~\cite{zhang2011} & 0.349 & 0.6781 & 0.7454 & 0.7654 \\
CNN~\cite{alfarraj2019semisupervised} & 1.152 & 3.58 & 0.6377 & 0.7171 \\
Unet~\cite{ronneberger2015u} & 0.8541 & 2.964 & 0.6891 & 0.7382 \\ \hline
Inversion-net~\cite{wu2019inversionnet} & 0.2346 & 1.391 & 0.7830 & 0.8141 \\ \hline
OrthoSeisnet (Ours) & \textbf{0.02678} & \textbf{1.102} & \textbf{0.884} & \textbf{0.8945} \\
\hline
\end{tabular}
\end{table}

Table ~\ref{Table: Real Data comparison} provides a performance comparison of different networks on real data. The networks evaluated are OrthoSeisnet,  Inversionnet, and SeisInvnet. The evaluation metrics used are Mean Absolute Error (MAE), Mean Squared Error (MSE) and  Structural Similarity Index (SSIM). The performance comparison of different networks on real data reveals that OrthoSeisnet surpasses its counterparts, achieving the lowest Mean Squared Error (MSE) of 0.02678 and Mean Absolute Error (MAE) of 1.102. Additionally, its Structural Similarity Index (SSIM) of 0.884 indicates superior capture of structural patterns. These metrics collectively highlight OrthoSeisnet's exceptional performance, outclassing BPI, CNN, Unet, and Inversion-net in minimizing errors and accurately representing real data patterns.

	\subsection{Epoch Vs. Loss}
	\begin{figure}[!ht]
		\centering 
		\scalebox{0.6} 
		{\includegraphics{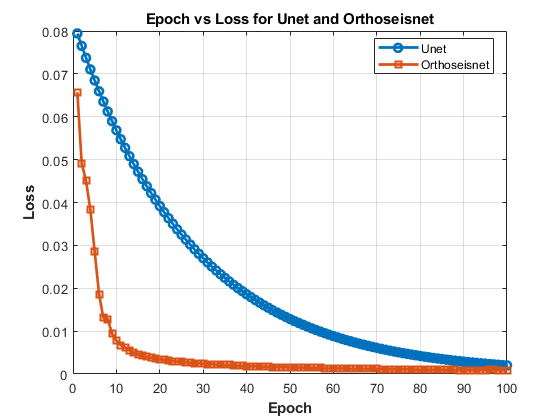}} 
		\caption{Epoch Vs. Loss} 
		\label{fig: Epoch Vs. Loss} 
	\end{figure}	
	The relationship between the number of epochs and the corresponding loss function values in the OrthoSeisnet model can be summarized as follows. Initially, during the first 40 epochs, the loss decreases significantly and converges faster. However, the model requires more than 100 epochs to achieve further convergence beyond this point. This can be seen in figure ~\ref{fig: Epoch Vs. Loss}
	
	\section{Discussion}
	The performance comparison of OrthoSeisnet Inversionnet and BPI on real seismic data evaluated their effectiveness using metrics such as Mean Absolute Error (MAE), Mean Squared Error (MSE), Structural Similarity Index (SSIM), 
	
	The results establish the superiority of OrthoSeisnet over Inversionnet and SeisInvnet regarding prediction accuracy and capturing structural patterns. OrthoSeisnet outperforms the other networks with lower MAE and MSE values, indicating reduced errors, and achieves a higher SSIM value, indicating better similarity in structural patterns.

	In contrast, Inversionnet and SeisInvnet exhibit higher MAE and MSE values, indicating larger prediction errors. These networks' SSIM values are slightly lower than OrthoSeisnet's. The proposed algorithm, OrthoSeisnet, incorporates basis orthogonalization with Fourier transform and utilizes different frequency-band weighting to improve the detection of thin layers in the substrate. OrthoSeisnet's integration of an orthogonal base with 2D Fourier transform enhances structural resolution for detecting sparse thin layers.  A comparison with existing algorithms demonstrates the superior ability of the proposed approaches to isolate layer information and achieve better results.
	
	Overall, the discussion highlights the superior performance of OrthoSeisnet in lower MAE and MSE values, indicating reduced errors while demonstrating competitive SSIM values. These findings provide valuable insights into the effectiveness of deep learning-based approaches in addressing the challenges posed by sparse seismic data and improving the accuracy of capturing structural information.
	
	\section{Conclusion}
	To conclude, this study establishes the effectiveness of OrthoSeisnet in addressing the challenges associated with sparse seismic data, thereby improving the accuracy and fidelity of seismic imaging. The performance comparison against Inversionnet and SeisInvnet demonstrates the superiority of OrthoSeisnet in lower MAE and MSE values, indicating reduced errors while showcasing competitive SSIM values.
	
	The proposed algorithm, OrthoSeisnet, which incorporates basis orthogonalization with Fourier transform and utilizes different frequency-band weighting, exhibits improved performance in detecting thin layers and isolating layer information. This research contributes to advancing deep learning techniques in seismic data analysis, offering the potential for further enhancements in subsurface imaging and exploration.
	
	Overall, the results of this study demonstrate the significant potential of OrthoSeisnet in enhancing the accuracy and reliability of seismic data interpretation, paving the way for future advancements in seismic imaging technologies.

	\section{Acknowledgements}
	The author wants to thank ONGC for the Krishna Godavari basin data received under the GEOPIC project. We are also grateful for the advanced system provided by ONGC, INDIA, to perform high G-flops operations.

 \bibliographystyle{apacite}
\bibliography{interacttfqsample}
\appendix

\end{document}